# Determination of the spin-lifetime anisotropy in graphene using oblique spin precession

Bart Raes[1], Jeroen E. Scheerder[2], Marius V. Costache[1], Frédéric Bonell[1], Juan F. Sierra[1], Jo Cuppens[1], Joris Van de Vondel[2] & Sergio O. Valenzuela[1,3]

We determine the spin-lifetime anisotropy of spin-polarized carriers in graphene. In contrast to prior approaches, our method does not require large out-of-plane magnetic fields and thus it is reliable for both low- and high-carrier densities. We first determine the in-plane spin lifetime by conventional spin precession measurements with magnetic fields perpendicular to the graphene plane. Then, to evaluate the out-of-plane spin lifetime, we implement spin precession measurements under oblique magnetic fields that generate an out-of-plane spin population. We find that the spin-lifetime anisotropy of graphene on silicon oxide is independent of carrier density and temperature down to 150 K, and much weaker than previously reported. Indeed, within the experimental uncertainty, the spin relaxation is isotropic. Altogether with the gate dependence of the spin lifetime, this indicates that the spin relaxation is driven by magnetic impurities or random spin-orbit or gauge fields.

[1] Catalan Institute of Nanoscience and Nanotechnology (ICN2), CSIC and The Barcelona Institute of Science and Technology, Campus UAB, Barcelona 08193, Spain. [2] INPAC—Institute for Nanoscale Physics and Chemistry, Department of Physics and Astronomy, KU Leuven, Celestijnenlaan 200D, B-3001 Leuven, Belgium. [3] Institució Catalana de Recerca i Estudis Avançats (ICREA), Barcelona 08070, Spain. Correspondence and requests for materials should be addressed to B.R. (email: bart.raes@icn.cat) or to S.O.V. (email: SOV@icrea.cat).

Identifying the main microscopic process for spin relaxation in graphene stands as one of the most fascinating puzzles for the graphene and spintronics communities[1–10]. Conventional Elliot–Yafet and Dyakonov–Perel relaxation mechanisms, which present opposite scalings between the spin lifetime $\tau_s$ and the momentum scattering time $\tau_p$, have often been utilized to gain insight into the spin-relaxation processes in graphene, but this type of analysis has yielded contradictory results[1,3]. For single-layer graphene, there are reports that are consistent with the Elliot–Yafet scaling, the Dyakonov–Perel scaling or even a combination of both[11–14]. For bilayer and multilayer graphene, the situation is similarly confusing[11,15,16]. To study the scaling, $\tau_p$ is tuned by either changing the carrier density $n$, by means of an external gate, or by introducing adatoms or molecules onto the graphene layer[17,18]. Nevertheless, this tuning will unavoidably lead to the modification of other parameters, relevant for spin transport. For example, in the case of spin-orbit-dominated spin relaxation, the nature and strength of the spin-orbit field can be energy (or charge density $n$) dependent, while the introduction of adatoms can actively mask or alter pre-existing spin-relaxation processes in an unintended way.

In this context, a fingerprint of the spin-orbit mechanism that can be determined as a function of $n$ and the type and density of the adatoms can provide valuable information about the induced modifications. The spin-lifetime anisotropy, which can be quantified by the ratio between the spin lifetimes for perpendicular and parallel spin components to the graphene plane $\zeta \equiv \tau_{s\perp}/\tau_{s\parallel}$, is precisely such a fingerprint. This anisotropy is determined by the preferential direction of the spin-orbit fields that cause the spin relaxation[1]. For spin-orbit fields preferentially in the graphene plane (for example, due to Rashba-type interaction), we expect $\zeta < 1$ (refs 5,19,20). For spin-orbit fields out-of-plane (for example, from ripples[5] or flexural distortions[21]), we expect $\zeta > 1$. There is no general relation for adatoms; depending on the element, they may induce spin-orbit fields with different directions and lead to specific signatures on the dependence of $\zeta$ with energy[22]. For example, in the case of fluorine adatoms, $\zeta$ is predicted to be equal to 0.5 for negative energies, but to depart markedly from this value for positive energies[23]. Finally, the presence of local magnetic moments with random orientation, originating from magnetic impurities or hydrogen adatoms, can overshadow the spin–orbit interaction, yielding $\zeta = 1$ (refs 24,25).

Despite their inherent interest, measurements of the spin-lifetime anisotropy are scarce[26,27]. Besides, the method to quantify the anisotropy requires intense out-of-plane magnetic fields $B_\perp > 1\,T$ (perpendicular to the graphene plane) and, therefore, it is expected to be useful for sufficiently large $n$ only, owing to the large magnetoresistive effects that are present in graphene at low carrier densities (see ref. 27 for further details). The actual value of $n$ beyond which the method would be suitable is sample dependent and, presumably, it increases with the mobility of graphene.

Here, we demonstrate an approach that overcomes the above limitation. The concept is based on spin precession measurements under oblique magnetic fields that generate an out-of-plane spin population, which is further used to evaluate the out-of-plane spin lifetime. The key of the method is to focus on the non-precessing spin component along the magnetic field, which will relax faster than the in-plane spin component, if $\zeta < 1$, or slower than the in-plane spin component, if $\zeta > 1$. We fabricate graphene devices with high mobility on silicon oxide substrates, however, the concept is general and can be implemented in any system that is susceptible to an anisotropic response. Our experiments demonstrate that the spin-relaxation anisotropy of graphene on silicon oxide is independent of carrier density and temperature down to 150 K, and much weaker than previously reported[26]. Altogether with the gate dependence of the spin lifetime, this indicates that the spin relaxation is driven either by random magnetic impurities[24,25] or by random spin-orbit fields or gauge fields[1,20]. These findings open the way for systematic anisotropy studies with tailored impurities and on different substrates. This information is crucial to find a route to increase the spin lifetime in graphene towards its intrinsic limit and, as such, has important implications for both fundamental science and technological applications.

## Results

**Measurement concept**. Our measurement scheme (Fig. 1) allows us to determine $\zeta$ at low magnetic fields ($B \sim 0.1\,T$), circumventing spurious magnetoresistive phenomena. In our approach, based on non-local spin devices, $\tau_{s\parallel}$ is first determined using conventional spin precession measurements, as shown schematically in Fig. 1a. This is performed by applying an out-of-plane magnetic field $B_\perp$, which causes the spins to precess exclusively in-plane as they diffuse from the injector (F1) towards the detector (F2). Because the magnetic field required to generate the spin precession is relatively weak (typically in the range of 0.1 T), the magnetizations of the injector and detector are assumed to remain in-plane over the whole-field range, due to magnetic shape anisotropy.

The measurement of $\tau_{s\perp}$ (Fig. 1b) relies on the fact that an oblique magnetic field[28], characterized by an angle $\beta$ (Fig. 1b, inset), causes the spins to precess out-of-plane as they diffuse towards F2. Therefore, the precession dynamics for $0 < \beta < 90°$ becomes sensitive to both $\tau_{s\parallel}$ and $\tau_{s\perp}$. As $B$ increases, the spin component perpendicular to the magnetic field dephases due to diffusive broadening. Eventually, for $B$ larger than the dephasing field $B_d$, only the component parallel to the field contributes to the non-local signal, which is picked up at the detector electrode. Dephasing greatly simplifies the data interpretation; the effective spin lifetime of the resulting component parallel to the field, $\tau_{s\beta}$, follows a simple relationship with $\zeta$ (see the 'Methods' section):

$$\frac{\tau_{s\beta}}{\tau_{s\parallel}} = \left(\cos^2(\beta) + \frac{1}{\zeta}\sin^2(\beta)\right)^{-1}. \quad (1)$$

By first measuring $\tau_{s\parallel}$ with the standard spin precession measurements (Fig. 1a) and then studying the dephased non-local signal as a function of $\beta$, we extract $\zeta$ using equation 1.

**Device design and characterization**. In our devices (Fig. 2a), we use graphene that is exfoliated onto a $p^{++}\,Si/SiO_2$ (440 nm) substrate from highly oriented pyrolytic graphite. Two inner ferromagnetic Co electrodes and two outer normal Pd electrodes contacting the graphene flake are fabricated using a single-electron-beam lithography step and shadow evaporation[29] (see Supplementary Fig. 1). Before metallization, an amorphous carbon layer is created between all contacts and the graphene flake by electron-beam overexposure of the contact area[30,31]. Shadow evaporation minimizes contamination from multiple lithographic steps. The amorphous carbon creates a resistive interface between the metals and the graphene (typically of about $10\,k\Omega$) that suppresses contact-induced spin relaxation and the conductivity mismatch problem[32], and that (associated with Co) is highly spin-polarized[30,33] (see the 'Methods' section for further details).

We first characterise the graphene charge transport properties by means of standard four-terminal local measurements, from which we estimate an average electron/hole mobility $\mu = 1.7 \times 10^4\,cm^2\,V^{-1}\,s^{-1}$ and a residual carrier density $n_0 = 1.5 \times 10^{11}\,cm^{-2}$, which sets the region where electron-hole puddles result in an inhomogeneous carrier density (see the 'Methods' section, Supplementary Fig. 2 and Supplementary Note 1). Figure 2b shows the normalized graphene resistance as a function of the strength of a perpendicular magnetic field $B_\perp$, in the low field range, for various gate voltages $V_g$ applied to the $p^{++}$ Si substrate. The magnetoresistance presents a peak at the charge-neutrality point (CNP)[34], $V_g = V_{CNP}$, but remains below 3% for all gate voltages when $B_\perp \le 0.2\,T$ (for further details see Supplementary Fig. 3 and Supplementary Note 2). The spin-transport properties in the graphene plane are then investigated using the non-local configuration, in which the current path is separated from the voltage detection circuit as shown in Fig. 2a. The current $I$ is driven between E1 and E2 and the non-local voltage $V_{nl}$ is measured between E3 and E4. Figure 2c shows typical non-local resistance $R_{nl} \equiv V_{nl}/I$ measurements as a function of the applied in-plane magnetic field. Sharp steps from positive-to-negative $R_{nl}$ are observed at $\pm 7$ and $\pm 15\,mT$, when the relative magnetization of the Co electrodes switches from parallel ($\downarrow\downarrow, \uparrow\uparrow$) to antiparallel ($\downarrow\uparrow$) configuration, or vice versa. Figure 3a shows the dependence of $\Delta R_{nl} \equiv R_{nl}^{\uparrow\uparrow} - R_{nl}^{\downarrow\uparrow}$ on $V_g$; the observed variation nearby the CNP is associated with a faster spin relaxation (see below) and a resistive, but not tunnelling, interface between the metal and graphene[4,35].

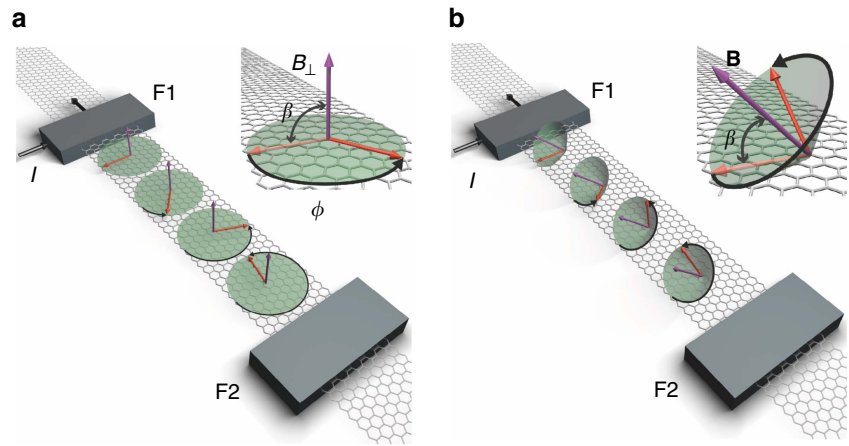

**Figure 1 | Spin precession and spin-lifetime anisotropy measurement principle.** (**a**) Conventional spin precession experiment in which the magnetic field $B_\perp$ is applied perpendicular to the graphene plane (purple arrow). A charge current (straight black arrows) through one of the ferromagnetic electrodes (F1) injects spins having an orientation parallel to the magnetization direction, which is fixed along the easy (that is, long) axis of the injector electrode. The injected spins (red arrows) undergo Larmor precession around $B_\perp$ while diffusing towards the detector electrode (F2). The precession angle $\phi$ changes with the strength of $B_\perp$, thus modulating the detected signal at F2. In this case, the precession is exclusively in the graphene plane and only sensitive to the parallel relaxation time (see inset). (**b**) Schematic illustration of the oblique-spin precession experiment proposed in this article. The magnetic field **B** is applied in a plane that contains the easy axis of the ferromagnetic electrodes and that is perpendicular to the substrate. For an oblique field, that is, $\beta \ne 0, 90°$ (see inset), the spins precess out-of-plane as they diffuse towards F2. In this situation, the effective spin lifetime is sensitive to both parallel and perpendicular spin lifetimes, $\tau_{s\|}$ and $\tau_{s\perp}$, and the spin-relaxation anisotropy can be experimentally obtained.

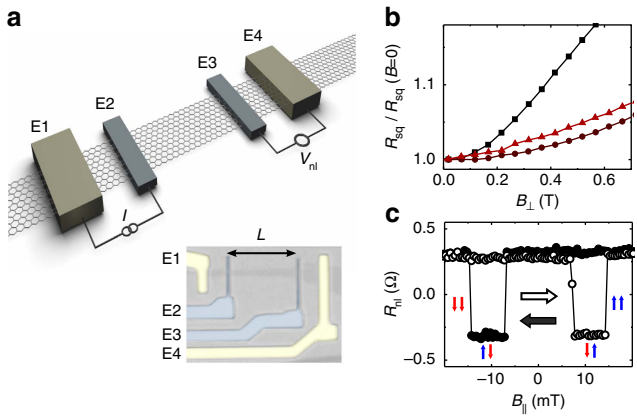

**Figure 2 | Device schematics and spin transport.** (**a**) Schematic drawing of the lateral non-local spin device geometry showing both the outer normal metallic electrodes (E1 and E4), and the inner ferromagnetic injector (E2) and detector (E3) electrodes. Wiring is shown in the non-local configuration, in which a current $I$ is applied between E1 and E2 and the non-local voltage $V_{nl}$ is measured between E3 and E4. An optical image of the non-local device used for the measurements shown in this paper is also shown. $L = 11\,\mu m$. (**b**) Normalized graphene square resistance $R_{sq}$ as a function of $B$. $V_g - V_{CNP} = 2.5, 42.5, 22.5\,V$ (top to bottom). The magnetoresistance is largest for $V_g - V_{CNP}$. (**c**) Non-local resistance $R_{nl} \equiv V_{nl}/I$ as a function of a magnetic field applied $B$ in-plane for up and down magnetic field sweeps at gate voltages $V_g$ such that $V_g - V_{CNP} = -25\,V$, with $V_{CNP}$ the position of the CNP. $T = 300\,K$ and $I = 10\,\mu A$.

**Conventional non-local spin precession.** A typical precession curve for magnetic fields perpendicular to the graphene plane is shown in Fig. 3b. We use these measurements to evaluate $\tau_{s\|}$ and the spin-diffusion coefficient $D_s$ as a function of $V_g$. To this end, we fit the data to the solution of the Bloch equations[4,36]. The results for $\tau_{s\|}$ and $D_s$ are presented in Fig. 3c, altogether with the spin-relaxation length $\lambda_{s\|} = \sqrt{\tau_{s\|} D_s}$. The spin-relaxation times and lengths in our device are as large as $\tau_{s\|} = 0.45\,ns$ and $\lambda_{s\|} = 5.8\,\mu m$, respectively, and increase away from the CNP. These values compare well with state-of-the-art studies of spin transport in non-encapsulated graphene on h-BN[14], and on suspended graphene[37,38].

To avoid magnetoresistive phenomena, shown in Fig. 2b, we note that complete dephasing of the spin component perpendicular to the magnetic field, at the position of the detector electrode, should occur at $B_d \lesssim 0.1\,T$. This condition is met at a large-enough injector-detector separation $L$. For typical graphene devices it is sufficient that $L \ge \sqrt{2}\lambda_s$, where $\lambda_s$ is the graphene spin-relaxation length (see Supplementary Note 3). The separation of the Co electrodes in the device of Fig. 2, $L = 11\,\mu m$, ensures that this condition is fulfilled, as $L$ is larger than the maximum value of $\sqrt{2}\lambda_s = 8.2\,\mu m$. As expected, Fig. 3b demonstrates that diffusive broadening completely suppresses the observation of spin precession at $B_d \sim 0.1\,T$. Using the fitting in Fig. 3b, we also evaluate the tilting angle $\gamma$ of the electrode magnetization out of the graphene plane (see Supplementary Fig. 4 and Supplementary Note 4). The results were verified with anisotropic magnetoresistance measurements in a Co wire (Supplementary Fig. 5 and Supplementary Note 5). We find that $\gamma$ remains below $\sim 5°$ for $B - B_d$, which guarantees that the extracted spin lifetimes are a reliable estimate for $\tau_{s\parallel}$. Finally, the data discussed so far was acquired at 300 K but the same conclusions apply to the measurements at 150 K. As observed in Fig. 3c, only weak variations are observed as the temperature is decreased.

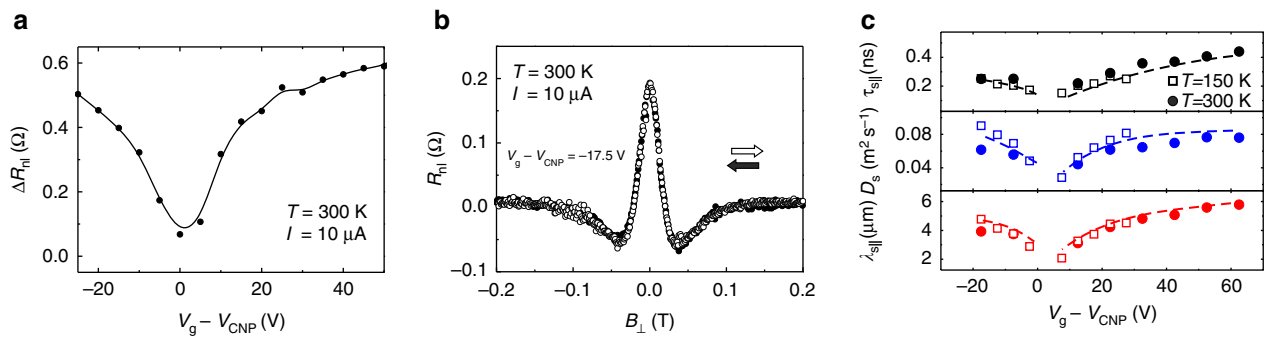

**Figure 3 | Spin precession under perpendicular magnetic field.** (a) Non-local resistance $R_{nl}$ change, $\Delta R_{nl} = R_{nl}^{\uparrow\uparrow} - R_{nl}^{\downarrow\uparrow}$, as a function of gate voltage $V_g$. The values of $\Delta R_{nl}$ are extracted from $R_{nl}$ versus in-plane magnetic field measurements and reflect the change in the signal when the relative magnetization of the ferromagnetic electrodes switches from parallel ($\downarrow\downarrow,\uparrow\uparrow$) to antiparallel ($\downarrow\uparrow$) configuration. The line is a guide to the eye. (b) Conventional spin precession measurements with perpendicular magnetic field $B_\perp$ for up (open symbols) and down (solid symbols) sweeps. Complete spin dephasing occurs for field strengths larger than $B_d \sim 0.1$ T. (c) In-plane spin-relaxation time $\tau_{s\parallel}$, spin-diffusion coefficient $D_s$ and spin-relaxation length $\lambda_{s\parallel} = \sqrt{\tau_{s\parallel} D_s}$, obtained from the spin precession measurements in b. The dashed lines are a guide to the eye. The measurements in a,b are performed in the non-local configuration. The error bars are smaller than the symbol size. The errors correspond to the standard error of mean obtained from the fitting for $D_s$ and $\lambda_{s\parallel}$, and the propagated error for $\tau_{s\parallel}$.

**Oblique non-local spin precession.** Having determined the in-plane spin-transport properties and demonstrated the high quality of the graphene sample, we proceed with the oblique-precession experiments to extract $\zeta$. Figures 4 and 5 present the main results of our work. Figure 4a shows spin precession measurements at $T = 300$ K for a representative set of $\beta$ values. As before, it is observed that the precessional motion dephases at $B_d \sim 0.1$ T. For $B > B_d$, $R_{nl}$ is determined by the remanent non-precessional spin component that lies along the magnetic field direction. Its magnitude, $R_{nl}^\beta \equiv R_{nl}(B > B_d)$, depends on $\beta$ and is nearly constant with increasing $B$. Figure 4b shows $R_{nl}^\beta$ versus $\beta$ at fixed $B = 175$ mT (marked by the vertical line in Fig. 4a) for the indicated values of $V_g$. For an isotropic system ($\zeta = 1$), $R_{nl}^\beta = R_{nl}^{\beta,\text{iso}} = R_{nl}^0 \cos^2(\beta^\star)$, where $R_{nl}^0 = R_{nl}(B=0)$ and $\beta^\star = \beta - \gamma(\beta, B)$ is the angle between the magnetization of the electrodes and the magnetic field, taking into consideration the small tilting $\gamma$. The factor $\cos^2(\beta^\star)$ accounts for the projection of the injected spins along the direction of the magnetic field and the subsequent projection along the direction of the detector magnetization.

**Determination of the spin-lifetime anisotropy.** For the general case of an anisotropic system, we determined $R_{nl}^\beta$ by solving the Bloch equations including both $\tau_{s\parallel}$ and $\tau_{s\perp}$, with $\zeta$ not necessarily equal to unity (see the 'Methods' section). We found that

$$R_{nl}^\beta = \sqrt{f(\zeta,\beta)} \exp\left[-\frac{L}{\lambda_\parallel}\left(\sqrt{\frac{1}{f(\zeta,\beta)}} - 1\right)\right] R_{nl}^{\beta,\text{iso}}, \quad (2)$$

where $f(\zeta,\beta) = \tau_{s\beta}/\tau_{s\parallel}$ is given by equation 1. For $\zeta = 1$, it is straightforward to verify that $R_{nl}^\beta = R_{nl}^{\beta,\text{iso}}$, regardless of the value of $\beta$.

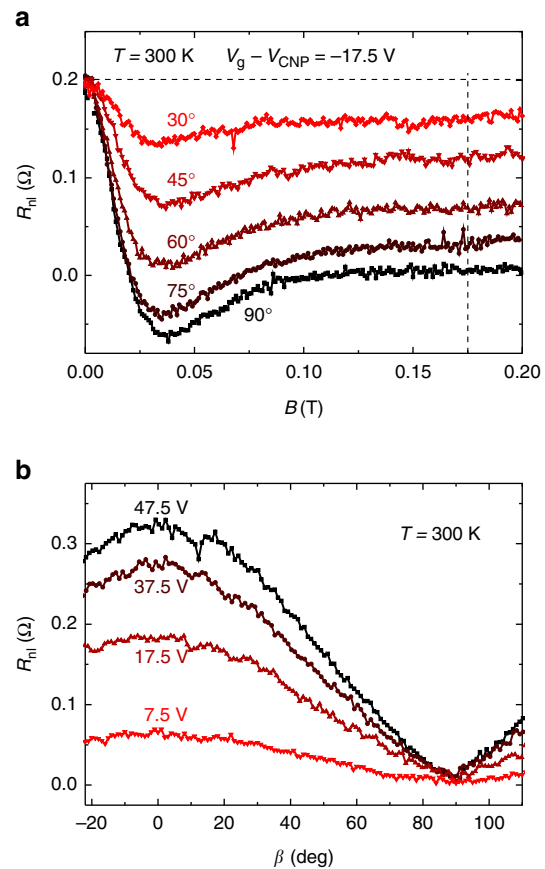

**Figure 4 | Spin precession measurements under oblique magnetic fields.** (a) Representative subset of experimental spin precession curves for $\beta = 30°, 45°, 60°, 75°, 90°$. The precession data are acquired at $T = 300$ K and $V_g - V_{CNP} = -17.5$ V, using an injector current of $I = 10$ $\mu$A, after preparing a parallel configuration of the electrodes. The horizontal dashed line is the non-local resistance at $B = 0$, $R_{nl}(B=0)$, which coincides with $R_{nl}$ at $\beta = 0°$ in the parallel configuration. (b) Angular dependence of $R_{nl}$ at fixed magnetic field $B = 175$ mT $> B_d$ for representative $V_g$. This magnetic field is shown with a vertical dashed line in a; $V_g - V_{CNP} = 47.5, 37.5, 17.5, 7.5$ V.

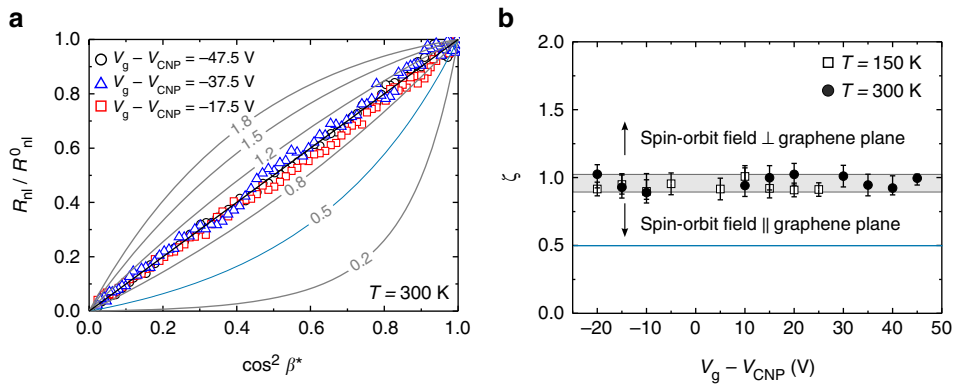

**Figure 5 | Spin-lifetime anisotropy ratio $\zeta$.** (**a**) Representative data from Fig. 4b normalized to 1, as a function of $\cos^2 \beta^*$, where $\beta^* = \beta - \gamma(\beta, B)$. The grey lines represent $R_{nl}^\beta$ for the indicated values of $\zeta$. The black straight line corresponds to $\zeta = 1$. (**b**) Extracted anisotropy ratio $\zeta$ as a function of $V_g$ for $T = 300$ K (solid circles) and $T = 150$ K (open squares). The blue solid lines in **a,b** mark the expected $\zeta = 0.5$ for in-plane spin-orbit fields. The grey area indicates $0.9 < \zeta < 1.03$. The error bars correspond to the standard error of mean from the fits in **a**. The main source of error is the noise of the measurements; propagated uncertainties in $\lambda_{s\|}$, $\beta$ and $\gamma$ produce a marginal contribution to the error in $\zeta$. The presented results are all obtained from the same sample.

Figure 5a shows the data in Fig. 4b normalized to 1, $R_{nl}^\beta/R_{nl}^0$, as a function of $\cos^2(\beta^*)$. The representation of Fig. 5a helps visualize any deviation from the isotropic case. As discussed above, for $\zeta = 1$, $R_{nl}^\beta = R_{nl}^0 \cos^2(\beta^*)$ and, therefore, $R_{nl}^\beta/R_{nl}^0$ versus $\cos^2(\beta^*)$ results in a straight line. According to equation 2, the response lies above the straight line for $\zeta > 1$ and below it for $\zeta < 1$. In Fig. 5a, we represent the predicted response for specific values of $\zeta$ from 0.2 to 1.8. We find that the experimental results (open symbols) are in excellent agreement with a straight line behaviour, thus $\zeta \sim 1$. The data is well-enclosed by the lines that correspond to $\zeta = 0.8$ and $\zeta = 1.2$, thereby defining rough lower and upper limits for $\zeta$. Figure 5b shows $\zeta$ as a function of $V_g$ for $T = 150$ K and 300 K by fitting the data to equation 2. All of the fitted values of $\zeta$ fall between $\sim 0.9$ and $\sim 1.03$ with no apparent dependence on either $V_g$ or $T$.

A comparison of the above results with those obtained with the methods shown in refs 26,27, can be achieved by applying a large perpendicular magnetic field that forces the magnetization direction of the electrodes to align to the field, therefore enabling the injection of spins perpendicular to the graphene plane. When complete rotation of the magnetization is obtained, $R_{nl}$ should saturate to a constant value from which $\tau_{s\perp}$ can be extracted. Unfortunately, we found that such a method could not be implemented in our sample, even for the largest carrier densities $n$ that we have studied. Indeed, although an apparent saturation of $R_{nl}$ is observed at $B \sim 1$ T, further increase of the magnetic field demonstrates that $R_{nl}$ is actually not independent of $B$, presenting a monotonous decrease, no matter the value of $n$. In the Supplementary Note 2 and Supplementary Fig. 3, we present the magnetoresistance effects in the graphene, while in Supplementary Note 6 and Supplementary Fig. 6 we discuss the results with perpendicularly magnetized electrodes and a criterion to evaluate the method applicability.

## Discussion

Considering the obtained anisotropy (Fig. 5b) and the gate dependence of $\tau_s$ and $D_s$ (Fig. 3c), we can draw several conclusions regarding the dominant mechanism of spin relaxation in our graphene samples. The temperature independence of all of the involved parameters demonstrates the small relevance of phonon scattering on the spin relaxation. In contradiction to the experiments, phonons are also predicted to lead to $\zeta > 1$ and to $\tau_s$ larger than tens of nanoseconds[1,21,39]. Because $\zeta \sim 1$, we can also rule out mechanisms associated with spin-orbit fields that are exclusively in-plane or out-of-plane of the graphene. Rashba spin–orbit coupling caused by (heavy) adatoms, such as gold, or the substrate, and the spin–orbit coupling caused by hydrogen-like adatoms (excluding magnetic moments)[23,40] predict a decreasing spin lifetime when approaching the CNP, as observed in the experiments (Fig. 3c). In the former, the faster relaxation at the CNP is due to a new spin-pseudospin entanglement mechanism specific to graphene[20,41]. In the latter, the faster relaxation originates from resonances close to the CNP that enhance the effect of the spin–orbit coupling[23]. However, in both cases $\zeta = 0.5$ should hold strictly, in disagreement with our results.

The anisotropy can be weaker for other adatoms that induce resonances, when the intrinsic spin–orbit coupling becomes important at off-resonance energies. Such is the case for fluor-like adatoms, although here $\zeta$ is expected to depend on energy and significantly deviate from 0.5 only at positive energies. In addition, spin-relaxation mechanisms associated with spin–orbit coupling from hydrogen- and fluorine-like adatoms predict $\tau_s$ to be larger than 100 ns for reasonable adatom coverage and therefore are unable to explain the ns-time scale observed in the experiments[23].

Besides Rashba-like spin-orbit fields, adatoms can introduce intrinsic-like spin-orbit fields, depending on the element under consideration. It is in principle possible to tailor any value of $\zeta$ for random distributions of specific adatoms. Moreover, increasing gauge fields, associated with strain, topological defects, or ripples can also result in a transition from $\zeta < 1$ to $\zeta > 1$ and mask a $\zeta = 0.5$ relation[5]. Their presence combined with spin-pseudospin entanglement could therefore lead to $\zeta \sim 1$ and, at the same time, explain the magnitude and carrier-density dependence of $\tau_s$. Although it appears unlikely to find exactly $\zeta = 1$ as in Fig. 5, experiments in graphene encapsulated with h-BN, which presumably reduces strain and contamination, are in favour of this interpretation[27]. In these experiments, the reported $\zeta$ is of the order of 0.7 at large $n$. However, measurements as a function of $n$ and at low magnetic fields are necessary to discard possible artefacts (Supplementary Note 6).

As discussed above, resonant scattering due to hydrogen-like adatoms can explain the observed dependence of $\tau_s$ on carrier density but not its magnitude. The reason is that, even though the resonances enhance the spin-flip scattering, the spin–orbit interaction is not effective because the resonance width (5 meV) is larger than the spin-orbit energy (1 meV) (ref. 23). Combining resonant scattering with local magnetic moments, which introduce the much larger exchange energy $|J| \sim 0.4$ eV, can lead to much shorter spin lifetimes while preserving the carrier-density dependence[24]. Calculations on hydrogenated graphene that capture the effect of magnetic moments predict spin lifetimes in the range of the experimental results with about 1 p.p.m. of hydrogen[24,25]. Because local magnetic moments are paramagnetic at such low concentrations, no preferential spin orientation is expected and thus $\zeta=1$. Within this model, chemisorbed hydrogen produces two peaks in $\tau_s^{-1}$: one above and one below the Dirac point with a separation of about 100 meV. To reproduce the experimental results, a smearing of the peaks of 110 meV is introduced, which is justified by the presence of charge fluctuations[24,25]. Spin anisotropy measurements at low temperatures in encapsulated graphene would therefore be the ultimate test for this model. Charge fluctuations should be much smaller in that case, in the range of 10 meV, allowing to resolve the two peaks and leading to a much sharper change in $\tau_s$ as a function of carrier density, and a maximum $\tau_s$ at the Dirac point, while $\zeta$ should remain equal to 1. These predictions could also be tested in bilayer graphene, where the energy broadening would be greatly reduced due to its larger density of states[42].

We have thus demonstrated spin-lifetime anisotropy measurements in graphene and discussed them in light of current theoretical knowledge. We used a measurement technique that provides reliable information on spin dynamics at low magnetic fields, and in a broad carrier-density range that was not accessible before. We show that only a very limited number of models can explain our results, and we provide a route based on our methods to discriminate between them using graphene spintronic devices that are within reach of the current state of the art. In addition, the microscopic properties of the graphene used in devices originating from different laboratories are not necessarily equivalent, owing to the graphite source or the processing steps that have been used. Therefore, it is plausible that experimental results in one research group might not be directly reproduced in another[1,11–14,17,18]. This underscores the importance of developing advanced spin-transport characterization techniques and the systematic implementation using samples of different origin. Spin-relaxation anisotropy measurements on specific substrates and with a controlled number of deposited adatoms will be crucial to increase the spin lifetime towards the theoretical limit, to find ways of controlling the spin lifetime, and to ultimately develop unprecedented approaches for the emergence of spin-based information processing protocols relying on graphene[2,43,44].

## Methods

**Device fabrication.** Graphene flakes are obtained by mechanically exfoliating highly oriented pyrolytic graphite (SPI Supplies) onto a p-doped Si substrate covered with 440 nm of $SiO_2$. The thickness of the $SiO_2$ layer is chosen to have optimum optical contrast between single- and multi-layer graphene flakes, which were discriminated by optical means after contrast calibration with Raman measurements. The width of the flake in the device shown in Fig. 2 is about $w \sim 1$ to 1.5 µm. An amorphous carbon (aC) interface is created between all contacts and the graphene flake by electron-beam (e-beam)-induced deposition before the fabrication of the contacts[30,31]. The aC deposition allows us to obtain large spin-signals by suppressing the conductivity mismatch problem, and the contact-induced spin relaxation (spin-sink effect); it is done by an e-beam overexposure of the contact area, with a dose about 30 times the typical working dose of e-beam resists.

We define the contact electrodes in our devices in a single e-beam lithography step using shadow evaporation to minimize processing contamination[29]. This results in devices with relatively large mobilities, which are in the range of $1 m^2 V^{-1} s^{-1}$ and thus compare well with similar devices on h-BN[14]. The shadow mask and evaporation steps are shown schematically in Supplementary Fig. 1. The mask is made from a 300-nm thick resist (MMA/PMMA) layer using MIBK:IPA (1:3) developer (the resists and the developer are from Microchem). All materials were deposited by e-beam evaporation in a chamber with a base pressure of $\sim 10^{-8}$ Torr. We first deposit the outer electrodes of Ti/Pd by angle deposition. Selecting an angle of ±45° from the normal to the substrate assures that no image of the lithographically produced lines reserved for the Co electrodes are deposited during the evaporation of Ti/Pd; indeed, Ti and Pd deposit onto the sidewalls of the lithography mask and are later on removed by lift-off[29,45]. Subsequently, Co is deposited under normal incidence. Co is in direct contact with the aC/graphene interface only for the inner electrodes, while for the outer electrodes the Ti/Pd layer is deposited in between. An undercut produced in the MMA layer assures that the deposited Co area on top of the outer Ti/Pd contacts is smaller by a factor of a few per cent. The nominal width $w_o$ and deposited thicknesses $d_{Ti}$ and $d_{Pd}$ of the identical outer Ti/Pd electrodes are $w_o = 1,500$ nm, and $d_{Ti} = 5$ nm and $d_{Pd} = 10$ nm. The inner Co contacts have widths $w_i^{E2} = 150$ nm and $w_i^{E3} = 140$ nm and thickness $d_{Co} = 30$ nm. The centre to centre distance between the Co contacts is $L = 11$ µm; their difference in width sets distinct coercive field strengths that allow us to control the relative orientation of their magnetization by an external magnetic field.

**Electrical characterization.** The devices are wired to a chip carrier that is placed in a variable temperature cryostat. Before start measuring, the sample space is first flushed with helium gas and then pumped, reaching an actual pressure of $<6.5 \times 10^{-4}$ Torr at 300 K and $1.5 \times 10^{-5}$ Torr at 150 K. We characterise the graphene charge transport properties by means of three- and four-terminal measurements. The contact resistances are of the order of 10 kΩ or larger, as determined by three-terminal measurements. The devices are homogeneous and we have not observed any significant shifts of the CNP in the different regions between each pair of contacts. For the four-terminal local measurements a current $I$ is driven between the outer Pd electrodes, E1 and E4, and the voltage measured between the inner Co electrodes, E2 and E3 (Fig. 2a). The characterization results are shown in Supplementary Fig. 2 and briefly discussed in Supplementary Note 1.

**Bloch-diffusion model including spin-lifetime anisotropy.** The contribution of the residual spin component parallel to the field, $s_{B_\parallel}$ ($s_\beta$ in the main text), to the non-local voltage is derived within the spin-diffusion model including spin anisotropy, which is quantified by the anisotropy ratio $\zeta$. In the isotropic limit ($\zeta = 1$), this model has been successfully used to describe spin precession phenomena in lateral devices[46], and in graphene when only the in-plane spin lifetime is relevant.

The spatio-temporal evolution of the spin density, $\mathbf{s}(x,t) = (s_x, s_{B_\parallel}, s_{B_\perp})$, on application of a homogeneous dc magnetic field $\mathbf{B} = (0, B, 0)$ can be described within a rotated cartesian axis system characterized by the unit vectors ($\hat{\mathbf{e}}_x, \hat{\mathbf{e}}_{B_\parallel}, \hat{\mathbf{e}}_{B_\perp}$) (see Supplementary Fig. 7) by,

$$\frac{\partial \mathbf{s}}{\partial t} = \overline{D_s} \nabla^2 \mathbf{s} + \gamma_c \mathbf{s} \times \mathbf{B} - \overline{\tau_s^{-1}} \cdot \mathbf{s}, \qquad (3)$$

where $\overline{D_s}$ is a scalar matrix with all diagonal entries equal to the spin-diffusion constant $D_s$, while $\overline{\tau_s^{-1}}$ is a symmetric (3 × 3) matrix describing spin relaxation with entries determined by the parallel and perpendicular spin lifetimes, $\tau_{s\parallel}$ and $\tau_{s\perp}$, and the angle $\beta$ of the applied magnetic field $\mathbf{B}$.

The injected spins feel a torque $\mathbf{N}$ due to the presence of the magnetic field, $\mathbf{N} = \gamma_c \mathbf{s} \times \mathbf{B}$, which results in a precessional evolution of the spin density. The constant pre-factor, $|\gamma_c| = g\mu_B/\hbar$, is the gyromagnetic ratio of the carriers, where $\mu_B$ is the Bohr magneton and $g$ is the so-called g-factor. In the case of exfoliated graphene, we can consider the g-factor to be equal to the free-electron value and to be field independent and isotropic.

In the limiting case in which the precessional motion is fully dephased, the spin density perpendicular to the field direction is suppressed, and the spatio-temporal evolution of $s_{B_\parallel}$, governed by equation 3, simplifies to,

$$\frac{\partial s_{B_\parallel}}{\partial t} = -\frac{s_{B_\parallel}}{\tau_{s\beta}} + D_s \frac{\partial^2 s_{B_\parallel}}{\partial x^2} \qquad (4)$$

with

$$\tau_{s_\beta}^{-1} = \frac{1}{\tau_{s\parallel}} \left[ \cos^2(\beta) + \frac{1}{\zeta} \sin^2(\beta) \right] = \frac{1}{\tau_{s\parallel}} f(\zeta, \beta)^{-1}. \qquad (5)$$

The above equation is equivalent to that governing the spin diffusion when only one spin lifetime is relevant, which here is given by the effective spin lifetime $\tau_{s_\beta}$ that is dependent on the direction of the field. The solution to this equation is well-known[36,46]; the non-local voltage $V_{nl}$ at the detector is,

$$V_{nl} = \alpha I \sqrt{\frac{\tau_{s_\beta}}{D_s}} e^{-\sqrt{\frac{L^2}{\tau_{s_\beta} D_s}}} [\cos(\beta - \gamma_i)][\cos(\beta - \gamma_d)], \qquad (6)$$

where $L$ indicates the distance between injector and detector and $I$ the magnitude of the injector current. The factor $\cos(\beta - \gamma_i)$ accounts for the projection of the injected spins along the direction of the magnetic field, corrected by a small tilting of the magnetization of the injector electrode, which is given by the angle $\gamma_i$. Similarly, the detector picks up the projection of the spin density along its own magnetization, whose orientation is corrected by an angle $\gamma_d$, leading to the $\cos(\beta - \gamma_d)$ factor (see also Supplementary Notes 4 and 5 and Supplementary Figs 4 and 5). The factor $\alpha$ depends on the square resistance of the graphene sheet and the effective polarization of the ferromagnetic electrodes. The effective polarization is a complex function, usually unknown, that depends on the materials and nature of the contacts involved (tunnelling, transparent, pinholes). Therefore, it is typically assumed to be a fitting parameter.

If we normalize the above result equation 6 to the value at $B=0$ taking into account equation 5 and considering $\gamma_i = \gamma_d = \gamma$, we obtain,

$$\frac{R_{nl}^\beta(B)}{R_{nl}(B=0)} = \frac{\alpha(B)}{\alpha(B=0)} \sqrt{f(\zeta, \beta)} e^{-\sqrt{\frac{L^2}{\tau_{s\parallel} D_s}} \left( \sqrt{f(\zeta, \beta)^{-1}} - 1 \right)} [\cos^2(\beta - \gamma)] \qquad (7)$$

where $R_{nl} = V_{nl}/I$.

At low-enough magnetic fields, the magnetoresistance can be disregarded and $\alpha(B)/\alpha(B=0) = 1$ (see Supplementary Note 2). By replacing $\lambda_\parallel^2 = \tau_\parallel D_s$, we obtain equation 2. Therefore, for the isotropic case, where $\zeta = 1$ and $f(\zeta, \beta) = 1$, the pre-factor in equation 7 is equal to one and the angular dependence of $R_{nl}$ simply follows $\cos^2(\beta - \gamma)$. In contrast, when $\zeta \neq 1$, the $\cos^2(\beta - \gamma)$ dependence is no longer valid. Because $\tau_{s\parallel}$ and $D_s$ can be independently determined from the conventional spin precession measurements ($\beta = 90°$), equation 7 provides a straightforward means of extracting the spin-lifetime anisotropy ratio $\zeta$ as a single fitting parameter from the experimental measurement of the angular dependence of $R_{nl}^{\beta}$.

## Acknowledgements

We thank A.W. Cummings and S. Roche for a critical reading of the manuscript and D. Torres for his help in developing Fig. 1. This research was partially supported by the European Research Council under Grant Agreement No. 308023 SPINBOUND, by the Spanish Ministry of Economy and Competitiveness, MINECO (under Contract No. MAT2013-46785-P and Severo Ochoa No. SEV-2013-0295), and by the Secretariat for Universities and Research, Knowledge Department of the Generalitat of Catalunya. M.V.C., J.F.S. and J.C. acknowledge support from the Ramón y Cajal, Juan de la Cierva and Beatriu de Pinós programs, respectively. F.B. acknowledges funding from the People Programme (Marie Curie Actions) of the European Union's Seventh Framework Programme FP7/2007-2013/ under REA Grant Agreement No. 624897. J.E.S. and J.V.d.V. acknowledge funding from the Methusalem Funding of the Flemish Government and the Research Foundation-Flanders (FWO).

## Author contributions

B.R. and S.O.V. planned the measurements and wrote the paper. B.R. fabricated the samples. J.E.S. and B.R. performed the measurements. B.R. modelled the results with input from all the authors. S.O.V. supervised the experiment. All authors commented on the manuscript.

## Additional information

**Supplementary Information** accompanies this paper at http://www.nature.com/naturecommunications

**Competing financial interests:** The authors declare no competing financial interests.

**Reprints and permission** information is available online at http://npg.nature.com/reprintsandpermissions/

**How to cite this article:** Raes, B. *et al.* Determination of the spin-lifetime anisotropy in graphene using oblique spin precession. *Nat. Commun.* 7:11444 doi: 10.1038/ncomms11444 (2016).